\documentclass[aps,manuscript,showpacs,preprintnumbers,amsmath,amssymb]{revtex4}
\usepackage[latin9]{inputenc}
\usepackage{units}
\usepackage{amsmath}
\usepackage{amssymb}

\makeatletter
\@ifundefined{textcolor}{}
{%
 \definecolor{BLACK}{gray}{0}
 \definecolor{WHITE}{gray}{1}
 \definecolor{RED}{rgb}{1,0,0}
 \definecolor{GREEN}{rgb}{0,1,0}
 \definecolor{BLUE}{rgb}{0,0,1}
 \definecolor{CYAN}{cmyk}{1,0,0,0}
 \definecolor{MAGENTA}{cmyk}{0,1,0,0}
 \definecolor{YELLOW}{cmyk}{0,0,1,0}
}

\usepackage{bm}
\usepackage{slashed}

\makeatother

\begin{document}

\title{New effects in the interaction between electromagnetic sources mediated
by nonminimal Lorentz violating interactions}

\author{L. H. C. Borges}
\email{luizhenriqueunifei@yahoo.com.br}

\affiliation{Universidade Federal do ABC, Centro de Ciências Naturais e Humanas,
Av. dos Estados, 5001, Santo André, SP, Brazil, 09210-580. }

\author{A. F. Ferrari}
\email{alysson.ferrari@ufabc.edu.br}

\affiliation{Universidade Federal do ABC, Centro de Ciências Naturais e Humanas,
Av. dos Estados, 5001, Santo André, SP, Brazil, 09210-580. }

\author{F. A. Barone}
\email{fbarone@unifei.edu.br}

\affiliation{IFQ - Universidade Federal de Itajubá, Av. BPS 1303, Pinheirinho,
Caixa Postal 50, 37500-903, Itajubá, MG, Brazil}
\begin{abstract}
This paper is dedicated to the study of interactions between external
sources for the electromagnetic field in the presence of Lorentz symmetry
breaking. We focus on a higher derivative, Lorentz violating interaction
that arises from a specific model that was argued to lead to interesting
effects in the low energy phenomenology of light pseudoscalars interacting
with photons. The kind of higher derivative Lorentz violating interaction
we discuss are called nonminimal. They are usually expected to be
relevant only at very high energies, but we argue they might also
induce relevant effects in low energy phenomena. Indeed, we show that
the Lorentz violating background considered by us leads to several
phenomena that have no counterpart in Maxwell theory, such as nontrivial
torques on isolated electric dipoles, as well as nontrivial forces
and torques between line currents and point like charges, as well
as among Dirac strings and other electromagnetic sources. 
\end{abstract}
\maketitle

\section{\label{I} Introduction}

The Standard Model (SM) of particle physics describes the fundamental
forces as well as the elementary particles that make up all matter,
being Lorentz and CPT invariant. However, in high energy scales of
the order of the Planck energy $E_{P}\sim10^{19}\mbox{GeV}$, it is
believed that quantum gravitational effects can not be neglected,
and there is the possibility of a spontaneous breaking of Lorentz
and CPT symmetries\,\cite{Planck}, or even a fundamental change
in the nature of quantum spacetime and its symmetries\,\cite{ACamelia1}.
In the last decades, the study of possible Lorentz symmetry violations
became an active field of theoretical and experimental research. The
motivation is essentially twofold: first, one hopes to learn from
eventual positive signs of Lorentz violation (LV) more on the fundamental
theory that operates at the Planck scale; second, from each negative
measure of LV one obtains a further test of Lorentz symmetry, leading
to an extensive set of contemporary, nontrivial tests of relativistic
symmetry\,\cite{DataTables}.

An early model to investigate the consequences of an explicit Lorentz
symmetry violation was proposed by Carroll, Field and Jackiw\,\cite{CFJ}.
In that work, the Maxwell Lagrangian was augmented by a kind of four
dimensional Chern-Simons term $k_{\mu}\epsilon^{\mu\nu\rho\sigma}A_{\nu}F_{\rho\sigma}$,
where the photon field couples to the Lorentz violating parameter
$k_{\mu}$. A result of this violation is a change in the propagation
of electromagnetic waves in the vacuum, which could be detected by
experiments, and whose absence induces an experimental constraint
on the LV parameter $k_{\mu}$.

A very systematic approach for the introduction of LV in the SM was
developed by Colladay and Kostelecký, in the form of the so-called
Standard Model Extension (SME). This model incorporates in the SM
structure all the Lorentz and CPT violating terms which respect renormalizability
and gauge invariance\,\cite{SME,SME1}. The SME constitutes a quite
general framework that facilitates investigations on the breaking
of Lorentz and CPT symmetries. Theoretical aspects of LV have been
investigated in Maxwell electrodynamics\,\cite{PRDqbk2004,EJPCxw2006,PRDbfho2003,PRDlp2004,PRLa2007},
QCD\,\cite{QCD}, gravity\,\cite{gravity,gravity1}, noncommutative
theories\,\cite{noncomutative}, statistical mechanics\,\cite{Mestatistical},
QED\,\cite{PLBck2001,PRDhlpw2009,QED,QED1}, supersymmetry\,\cite{SUSY,SUSY1,SUSY11,SUSY2},
electromagnetic wave propagation\,\cite{ABBGHplb2012,CFMepjc2012,ra1},
to name a few. Experimental tests of LV have been performed in experiments
involving photons\,\cite{photon1,photon2}, electrons\,\cite{electron1,electron2},
muons\,\cite{muons1,muons2}, and many others\,\cite{DataTables}.

The SME, understood as an effective field theory, includes renormalizable
LV interactions as well as higher derivatives, non renormalizable
ones. These later are called nonminimal terms, and by dimensional
analysis alone are expected to be subdominant relative to the minimal
ones, so in principle they could be disregarded except if one deals
with extremely high energy phenomena. The systematic study of the
nonminimal terms of the SME was started in\,\cite{hiphsec,hifesec},
where higher derivatives terms in both photon and fermion sectors
were considered. Besides, some studies of different nonminimal LV
interactions were carried out in\,\cite{hilv2,hilv1,hilv4,hilv5,hilv6,ABBGHplb2012,hilv3},
for example.

In\,\cite{ALP}, it was shown that nonminimal terms can induce nontrivial
effects even in low energy phenomenology, in particular in searches
for light pseudoscalars. The QCD axion being its most important representative,
these kind of particles are extremely light and weakly interacting,
thus constituting a class of WISPs (weakly interacting, slim particles)
candidates for dark matter\,\cite{Jaeckel:2010ni,Ringwald:2012hr}.
A particular set of nonminimal LV interactions was shown to provide
a mechanism for generating the standard, Lorentz invariant (LI), interaction
between the photon and light pseudoscalars that is investigated by
current experimental efforts. Despite the nonminimal LV interactions
being assumedly very small, they might represent a relevant contribution
to these phenomenology, since the standard LI interactions involving
WISPs are themselves very feeble.

The lesson is that there might exist open windows for the investigation
of nonminimal LV effects in low energy physics, and photon physics
is a natural place to start looking for this window. A starting point
in this direction was the work\,\cite{Highoperators}, in which the
complete low energy photon effective action for the nomminimal LV
interaction studied in\,\cite{ALP} was calculated. This result paves
the way to investigate possible nonminimal LV effects in electrodynamics.
A first result, already discovered in\,\cite{Highoperators}, is
that the propagation of electromagnetic waves in the vacuum is not
affected by the particular LV coupling considered in these articles.

In the present work, we look further for nontrivial effects of nonminimal
LV in the classical interaction between electromagnetic sources. Our
analysis parallels that of reference\,\cite{Fontes}, which considered
a minimal LV term from the non birefringent sector of the SME. As
in that paper, we will find essential modifications due to the LV,
with some effects that have no counterparts in the standard Maxwell
theory. These will be results obtained without the recourse to perturbation
theory.

The paper is organized as follows: in section\,\ref{II} we define
the specific model we will study, and calculate the (exact) photon
propagator. This result is used to obtain the classical interaction
between different electromagnetic sources: point-like stationary charges
(section\,\ref{III}), a steady line current and a point-like stationary
charge (section\,\ref{IV}), and Dirac strings (section\,\ref{V}).
Finally, section\,\ref{conclusoes} is dedicated to our final remarks
and conclusions. Along the paper we shall deal with models in $3+1$
dimensional space-time and use Minkowski coordinates with the diagonal
metric with signature $(+,-,-,-)$.

\section{\label{II}The model}

The explicit model we will consider in this work is defined by the
following Lagrangian density, 
\begin{align}
{\cal L}= & -\frac{1}{4}F_{\mu\nu}F^{\mu\nu}-\frac{1}{2\gamma}\left(\partial_{\mu}A^{\mu}\right)^{2}+\frac{1}{2}d^{\lambda}d_{\alpha}\partial_{\mu}F_{\nu\lambda}\partial^{\nu}F^{\mu\alpha}\nonumber \\
 & +J^{\mu}A_{\mu}\ ,\label{eq:1}
\end{align}
where $A^{\mu}$ is the electromagnetic field, $F^{\mu\nu}=\partial^{\mu}A^{\nu}-\partial^{\nu}A^{\mu}$
is the field strength, $J^{\mu}$ is the external source, $\gamma$
is a gauge parameter, and $d^{\lambda}$ is a background vector taken
to be constant and uniform in the reference frame where the calculations
are performed. The parameter $d^{\lambda}$ embodies the LV in our
model. We restrict ourselves to the case of $d^{\mu}$ being a time-like
background vector, namely $d^{2}=d^{\mu}d_{\mu}>0$. The case of a
light-like background vector can be obtained from our results by taking
the limit $d^{2}\to0$, in which case all effects due to the LV disappear.
The case of a space-like background vector is more subtle. Some preliminary
results suggest that the interaction energy could exhibit an imaginary
part in some circumstances, making the vacuum unstable in the presence
of external sources; in addition, we could have tachyonic modes. These
facts can be an indication that the model is not consistent for a
space-like background vector, so this situation will not be considered
in this paper.

Differently from the minimal LV coefficients appearing in the SME,
the parameter $d^{\lambda}$ is not adimentional but instead has length
dimension one. The LV term in\,\eqref{eq:1} is one of the low energy
interactions obtained in\,\cite{Highoperators}, starting from the
basic LV interaction $F_{\mu\nu}d^{\nu}\overline{\psi}\gamma^{\mu}\psi$,
between the photon and a very massive fermion field $\psi$ which
is integrated out to study the low energy phenomenology of the model.

The propagator $D^{\mu\nu}(x,y)$ for the Lagrangian \eqref{eq:1}
satisfies the differential equation 
\begin{align}
\Biggl\{\partial^{2}\eta^{\mu\nu}-\Biggl[\left(1-\frac{1}{\gamma}\right)-\left(d\cdot\partial\right)^{2}\Biggr]\partial^{\mu}\partial^{\nu}+d^{\mu}d^{\nu}\partial^{4}\nonumber \\
-\partial^{2}\left(d\cdot\partial\right)\left(d^{\mu}\partial^{\nu}+d^{\nu}\partial^{\mu}\right)\Biggr\}{D_{\nu}}^{\beta}\left(x,y\right)\nonumber \\
=\eta^{\mu\beta}\delta^{4}\left(x-y\right)\ .
\end{align}
Fixing the Feynman gauge $\gamma=1$, one can solve this equation
obtaining the exact propagator in the form of the Fourier integral
\begin{align}
D^{\mu\nu}(x,y) & =\int\frac{d^{4}p}{(2\pi)^{4}}\Biggl\{-\frac{\eta^{\mu\nu}}{p^{2}}+\frac{1}{[1-d^{2}p^{2}+(p\cdot d)^{2}]}\nonumber \\
 & \times\Biggl[-d^{\mu}d^{\nu}-\frac{(p\cdot d)^{2}}{p^{4}}p^{\mu}p^{\nu}\nonumber \\
 & +\frac{(p\cdot d)}{p^{2}}(p^{\mu}d^{\nu}+d^{\mu}p^{\nu})\Biggr]\Biggr\} e^{-ip\cdot(x-y)}\ .\label{eq:propagator}
\end{align}
This propagator is the basic ingredient we need to obtain several
relevant physical quantities of the model. We shall use it to compute
the interaction mediated by the electromagnetic field between several
different sources in the next sections.

\section{\label{III}Point-like charges}

In this section we consider the interaction between two steady point-like
charges in the model defined by Eq.\,\eqref{eq:1}. This charge configuration
is described by the external source 
\begin{equation}
J_{\mu}^{I}({\bf x})=q_{1}\eta_{\mu}^{0}\delta^{3}\left({\bf x}-{\bf a}_{1}\right)+q_{2}\eta_{\mu}^{0}\delta^{3}\left({\bf x}-{\bf a}_{2}\right)\ ,\label{corre1Em}
\end{equation}
where the location of the charges are specified by the vectors ${\bf a}_{1}$
and ${\bf a}_{2}$. The parameters $q_{1}$ and $q_{2}$ are the coupling
constants between the vector field and the delta functions, and can
be interpreted as electric charges.

The theory is quadratic in the field variables $A^{\mu}$, so it can
be shown that the contribution of the source $J(x)$ to the vacuum
energy of the system is given by\,\cite{Zee,BaroneHidalgo1,BaroneHidalgo2}
\begin{equation}
E=\frac{1}{2T}\int d^{4}y\int d^{4}x\ J_{\mu}(x)D^{\mu\nu}(x,y)J_{\nu}(y)\ ,\label{zxc1}
\end{equation}
where the integration in $y^{0}$ is from $-\nicefrac{T}{2}$ to $\nicefrac{T}{2}$,
and the limit $T\to\infty$ is implicit.

Substituting (\ref{corre1Em}) into\,(\ref{zxc1}), discarding the
self-interacting energy of each charge, we have 
\begin{align}
E^{I}= & \frac{q_{1}q_{2}}{T}\int d^{4}y\int d^{4}x\ d^{4}y\ D^{00}(x,y)\nonumber \\
 & \times\delta^{3}\left({\bf x}-{\bf a}_{1}\right)\delta^{3}\left({\bf y}-{\bf a}_{2}\right)\ .
\end{align}
By using the explicit form of the propagator in Eq.\,(\ref{eq:propagator}),
computing the integrals in the following order: $d^{3}{\bf x}$, $d^{3}{\bf y}$,
$dx^{0}$, $dp^{0}$ and $dy^{0}$, using the Fourier representation
for the Dirac delta function $\delta(p^{0})=\int dx/(2\pi)\exp(-ipx)$,
and identifying the time interval as $T=\int_{-\frac{T}{2}}^{\frac{T}{2}}dy^{0}$,
we can write 
\begin{eqnarray}
E^{I} & = & q_{1}q_{2}\int\frac{d^{3}{\bf p}}{(2\pi)^{3}}\frac{\exp(i{\bf p}\cdot{\bf a})}{{\bf p}^{2}}\nonumber \\
 & - & \frac{q_{1}q_{2}(d^{0})^{2}}{d^{2}}\int\frac{d^{3}{\bf p}}{(2\pi)^{3}}\frac{\exp(i{\bf p}\cdot{\bf a})}{\left({\bf p}^{2}+\frac{\left({\bf d}\cdot{\bf p}\right)^{2}}{d^{2}}\right)+\frac{1}{d^{2}}}\ ,\label{Ener2EM}
\end{eqnarray}
where $d=\sqrt{d^{2}}$ and we defined ${\bf a}={\bf a}_{1}-{\bf a}_{2}$,
which is the distance between the two electric charges. Remembering
that 
\begin{eqnarray}
\int\frac{d^{3}{\bf p}}{(2\pi)^{3}}\frac{\exp(i{\bf p}\cdot{\bf a})}{{\bf p}^{2}}=\frac{1}{4\pi|{\bf a}|}\ ,\label{Ener4EM1}
\end{eqnarray}
we can note that the first term in Eq. (\ref{Ener2EM}) gives the
well-known Coulombian interaction.

In order to calculate the second integral in Eq.\,(\ref{Ener2EM}),
we shall perform a change in the integration variables. For this task,
we first split the vector ${\bf p}$ into two parts, 
\begin{equation}
{\bf p}={\bf p}_{n}+{\bf p}_{p}\thinspace,
\end{equation}
${\bf p}_{p}$ being parallel and ${\bf p}_{n}$ normal to the vector
${\bf d}$; more explicitly, 
\begin{equation}
{\bf p}_{p}={\bf d}\Bigl(\frac{{\bf d}\cdot{\bf p}}{{\bf d}^{2}}\Bigr),\ \ {\bf p}_{n}={\bf p}-{\bf d}\Bigl(\frac{{\bf d}\cdot{\bf p}}{{\bf d}^{2}}\Bigr)\thinspace.\label{mudan1EM}
\end{equation}
We also define the vector ${\bf q}$ as follows, 
\begin{eqnarray}
{\bf q} & = & {\bf p}_{n}+{\bf p}_{p}\sqrt{1+\frac{{\bf d}^{2}}{d^{2}}}\label{defq}\\
 & = & {\bf p}+{\bf d}\Bigl(\frac{{\bf d}\cdot{\bf p}}{{\bf d}^{2}}\Bigr)\left(\frac{\left|d^{0}\right|}{d}-1\right)\ .
\end{eqnarray}
With the previous definitions, we can write 
\begin{equation}
{\bf p}_{p}=\frac{{\bf d({\bf d\cdot{\bf q)}}}}{{\bf d}^{2}}\frac{d}{|d^{0}|}\ ,\ {\bf p}_{n}={\bf q-\frac{{\bf d({\bf d\cdot{\bf q)}}}}{{\bf d}^{2}}}\thinspace,\label{mudan6EM}
\end{equation}
which implies in 
\begin{equation}
{\bf p}={\bf q}+\frac{({\bf {d}\cdot{\bf q}){\bf {d}}}}{{\bf d}^{2}}\left(\frac{d}{|d^{0}|}-1\right)\thinspace,
\end{equation}
and 
\begin{equation}
{\bf q}^{2}={\bf p}^{2}+\frac{({\bf d}\cdot{\bf p})^{2}}{d^{2}}\ .\label{zxc2}
\end{equation}
Defining the spatial vector 
\begin{equation}
{\bf b}={\bf a}+\left(\frac{d}{|d^{0}|}-1\right)\frac{{\bf d}\cdot{\bf a}}{{\bf d}^{2}}{\bf d}\ ,\label{zxc3}
\end{equation}
and using Eq.\,(\ref{mudan6EM}), we can show that 
\begin{equation}
{\bf p}\cdot{\bf a}={\bf b}\cdot{\bf q}\ .\label{mudan3EM}
\end{equation}
The Jacobian of the transformation from ${\bf p}$ to ${\bf q}$ can
be calculated from Eq.\,(\ref{mudan6EM}), resulting in 
\begin{equation}
\det\left[\frac{\partial{\bf {p}}}{\partial{\bf {q}}}\right]=\frac{1}{\sqrt{1+\frac{{\bf d}^{2}}{d^{2}}}}=\frac{d}{|d^{0}|}\ .\label{mudan5EM}
\end{equation}
Putting all this together, we end up with 
\begin{equation}
E^{I}=\frac{q_{1}q_{2}}{4\pi|{\bf a}|}-\frac{q_{1}q_{2}|d^{0}|}{d}\int\frac{d^{3}{\bf q}}{(2\pi)^{3}}\frac{\exp(i{\bf b}\cdot{\bf q})}{{\bf q}^{2}+\frac{1}{d^{2}}}\ .\label{Ener3EM}
\end{equation}

Using the fact that, for $d^{2}>0$\,\cite{BaroneHidalgo1}, 
\begin{equation}
\int\frac{d^{3}{\bf q}}{(2\pi)^{3}}\frac{\exp(i{\bf b}\cdot{\bf q})}{{\bf q}^{2}+\left(\frac{1}{d^{2}}\right)}=\frac{1}{4\pi|{\bf b|}}\exp\left(-\frac{|{\bf b|}}{d}\right)\ ,\label{Ener4EM}
\end{equation}
and performing some manipulations we arrive at, 
\begin{equation}
E^{I}=\frac{q_{1}q_{2}}{4\pi}\Biggl[\frac{1}{|{\bf a}|}-\frac{|d^{0}|}{d}\frac{1}{|{\bf b}|}\exp\Biggl(-\frac{|{\bf b}|}{d}\Biggr)\Biggr]\ ,\label{Ener6EM}
\end{equation}
where, 
\begin{equation}
|{\bf b}|=\sqrt{{\bf a}^{2}-\frac{\left({\bf d}\cdot{\bf a}\right)^{2}}{|d^{0}|^{2}}}\ .\label{Ener66EM}
\end{equation}
It is important to realize that $|{\bf b}|$ vanishes only of ${\bf a}=0$.
This can be seen by taking a coordinate system where ${\bf d}$ lies
along the ${\hat{z}}$ axis. In spherical coordinates, with $\theta$
standing for the polar angle for ${\bf a}$, Eq.\,(\ref{Ener66EM})
reads 
\begin{eqnarray}
|{\bf b}|=|{\bf a}|\sqrt{1-\Biggl(\frac{|{\bf d}|}{|d^{0}|}\cos(\theta)\Biggr)^{2}}\ .\label{rfv1}
\end{eqnarray}
The restriction $d^{2}>0$ guarantees that the term inside the square
root will always be strictly positive.

Equation\,(\ref{Ener6EM}) gives the interaction energy between two point-like charges mediated by
the electromagnetic field with the specific Lorentz violating coupling
contained in Eq.\,\eqref{eq:1}. The $d^{\mu}$ dependent term in\,(\ref{Ener6EM})
is a correction to the Coulomb interaction due the Lorentz symmetry
breaking, leading to an anisotropic interaction between the charges.
The LI limit $d^{\mu}\to0$ of this result must be taken with the
restriction $(d^{0})^{2}>{\bf d}^{2}$: in this case, it can shown
that the second term inside the brackets vanishes and we are left
with the standard Coulombian interaction between the charges. The
same happens for the light-like limit of the vector $d^{\mu}$, i.e.,
if $d\rightarrow0$.

If ${\bf d}=0$, Eq. (\ref{Ener6EM}) reduces to the simple form 
\begin{equation}
E^{I}({\bf d}=0)=\frac{q_{1}q_{2}}{4\pi}\Biggl[\frac{1}{|{\bf a}|}-\frac{1}{|{\bf a}|}\exp\Biggl(-\frac{|{\bf a}|}{|d^{0}|}\Biggr)\Biggr]\ ,\label{rfv2}
\end{equation}
which is the Coulombian interaction corrected by an Yukawa-like interaction,
with $1/|d^{0}|$ as a mass parameter. It is interesting to notice
that the same structure for $E^{I}$ can be found for another (Lorentz
invariant) gauge field theory which exhibits higher order derivatives\,\cite{BNH1},
the Podolsky-Lee-Wick electrodynamics\,\cite{Podolsky,LW}. As another
noteworthy particular case, if the distance vector ${\bf a}$ is perpendicular
to the background vector ${\bf d}$, Eq.\,(\ref{Ener6EM}) becomes
\begin{equation}
E^{I}({\bf d}\cdot{\bf a}=0)=\frac{q_{1}q_{2}}{4\pi}\Biggl[\frac{1}{|{\bf a}|}-\frac{|d^{0}|}{d}\frac{1}{|{\bf a}|}\exp\Biggl(-\frac{|{\bf a}|}{|d^{0}|}\Biggr)\Biggr]\ .
\end{equation}
The force between the two charges can be calculated from Eqs.\,(\ref{Ener6EM})
and\,(\ref{Ener66EM}), resulting in 
\begin{align}
{\bf F}^{I} & =-\nabla E^{I}\nonumber \\
 & =\frac{q_{1}q_{2}}{4\pi}\Biggl[\frac{{\bf a}}{|{\bf a}|^{3}}-\frac{|d^{0}|}{d}\frac{1}{|{\bf b}|^{3}}\Biggl(1+\frac{|{\bf b}|}{d}\Biggr)\nonumber \\
 & \ \ \ \times\exp\Biggl(-\frac{|{\bf b}|}{d}\Biggr)\Biggl({\bf a}-\frac{\left({\bf d}\cdot{\bf a}\right){\bf d}}{|d^{0}|^{2}}\Biggr)\Biggr]\thinspace.
\end{align}

The interaction energy (\ref{Ener6EM}) exhibits anisotropy due to
the presence of the background vector $d^{\mu}$. An interesting consequence
of this is the emergence of an spontaneous torque on an electric dipole.
To see this, we consider a typical dipole composed by two opposite
electric charges, $q_{1}=-q_{2}=q$, placed at the positions ${\bf a}_{1}={\bf R}+\frac{{\bf A}}{2}$
and ${\bf a}_{2}={\bf R}-\frac{{\bf A}}{2}$, ${\bf A}$ taken to
be a fixed vector. From Eq.\,(\ref{Ener6EM}), we obtain 
\begin{align}
E_{dipole}=-\frac{q^{2}}{4\pi|{\bf A|}}\Biggl[1-\frac{|d^{0}|}{d}\frac{1}{f(\Theta)}\exp\Biggl(-\frac{|{\bf A}|f(\Theta)}{d}\Biggr)\Biggr]\thinspace,
\end{align}
where 
\begin{equation}
f(\Theta)=\sqrt{1-\frac{{\bf d}^{2}\cos^{2}\Theta}{|d^{0}|^{2}}}\ ,\label{modb1EMMMM}
\end{equation}
with $\Theta\in\left[0,2\pi\right)$ standing for the angle between
${\bf A}$ and the background vector ${\bf d}$. This interaction
energy leads to an spontaneous torque on the dipole as follows, 
\begin{align}
\tau_{dipole}= & -\frac{\partial E_{dipole}}{\partial\Theta}\nonumber \\
= & \frac{q^{2}}{8\pi|{\bf A|}}\frac{{\bf d}^{2}}{d|d^{0}|}\frac{1}{f^{3}(\Theta)}\Biggl(1+\frac{|{\bf A}|f(\Theta)}{d}\Biggr)\nonumber \\
 & \ \times\sin(2\Theta)\exp\Biggl(-\frac{|{\bf A}|f(\Theta)}{d}\Biggr)\ .
\end{align}
This spontaneous torque on the dipole is an exclusive effect due to
the Lorentz violating background. If $d^{\mu}=0$, the torque vanishes,
as it should, as well as for the specific configurations $\Theta=0,\pi/2,\pi$.
Finally, we note that if ${\bf d}=0$, this effect is also absent.

\section{\label{IV}A steady current line and a point-like charge}

In this section we study the interaction energy between a steady line
current and a point-like stationary charge. This interaction does
not occur in Maxwell electrodynamics, but it may emerge in theories
with Lorentz violation\,\cite{Fontes}, as well as in LI theories
with higher order derivatives\,\cite{BNH1}.

Let us consider a steady line current flowing parallel to the $z$-axis,
along the straight line placed at ${\bf A}=(A^{1},A^{2},0)$. The
electric charge is placed at the position ${\bf s}$. The external
source for this system is given by 
\begin{equation}
J_{\mu}^{II}\left({\bf x}\right)=I\eta_{\ \mu}^{3}\delta^{2}\left({\bf x}_{\perp}-{\bf A}\right)+q\eta_{\ \mu}^{0}\delta^{3}\left({\bf x}-{\bf s}\right)\ ,\label{corre3Em}
\end{equation}
where we defined the vector position perpendicular to the straight
line current ${\bf x}_{\perp}=(x^{1},x^{2},0)$. The parameters $I$
and $q$ stand for, respectively, the current intensity and the electric
charge.

Substituting\,(\ref{corre3Em}) into\,(\ref{zxc1}) and discarding
self-interaction terms, we have 
\begin{align}
E^{II}= & \frac{qI}{T}\int d^{4}y\int d^{4}x\ D^{30}(x,y)\nonumber \\
 & \times\delta^{2}\left({\bf x}_{\perp}-{\bf A}\right)\delta^{3}\left({\bf y}-{\bf s}\right)\ ,
\end{align}
where the integration limits for $y^{0}$ are as in the previous section.
Substituting the explicit form for the propagator\,(\ref{eq:propagator})
and evaluating the integrals $d^{2}{\bf x}_{\perp}$, $d^{3}{\bf y}$,
$dx^{3}$, $dp^{3}$, $dx^{0}$, $dp^{0}$ and $dy^{0}$, we obtain
\begin{equation}
E^{II}=-\frac{qId^{3}d^{0}}{d^{2}}\int\frac{d^{2}{\bf p}_{\perp}}{(2\pi)^{2}}\frac{\exp(i{\bf p}_{\perp}\cdot{\bf a}_{\perp})}{\left({\bf p}_{\perp}^{2}+\frac{\left({\bf d}_{\perp}\cdot{\bf p}_{\perp}\right)^{2}}{d^{2}}\right)+\frac{1}{d^{2}}}\thinspace,\label{Ener9EM}
\end{equation}
where again $\int_{-T/2}^{T/2}dy^{0}=T$, and defined the perpendicular
momentum ${\bf p}_{\perp}=(p^{1},p^{2},0)$ and the distance between
the charge and the line current ${\bf a}_{\perp}=(A^{1}-s^{1},A^{2}-s^{2},0)$.

Proceeding as in before, the interaction energy in this case can be
written as 
\begin{equation}
E^{II}=-\frac{qId^{3}d^{0}}{d\sqrt{(d^{0})^{2}-(d^{3})^{2}}}\int\frac{d^{2}{\bf q}_{\perp}}{(2\pi)^{2}}\frac{\exp(i{\bf q}_{\perp}\cdot{\bf r}_{\perp})}{{\bf q}_{\perp}^{2}+\frac{1}{d^{2}}}\thinspace,\label{Ener10EMM}
\end{equation}
where we defined 
\begin{equation}
{\bf r}_{\perp}={\bf a}_{\perp}+\Biggl[\frac{d}{\sqrt{(d^{0})^{2}-(d^{3})^{2}}}-1\Biggr]\frac{{\bf d}_{\perp}\cdot{\bf a}_{\perp}}{{\bf d}_{\perp}^{2}}{\bf d}_{\perp}\ .\label{defr}
\end{equation}
Due to the fact that $d^{2}>0$, we can use that\,\cite{BaroneHidalgo1}
\begin{equation}
\int\frac{d^{2}{\bf q}_{\perp}}{(2\pi)^{2}}\frac{\exp(i{\bf q}_{\perp}\cdot{\bf r}_{\perp})}{{\bf q}_{\perp}^{2}+\left(\frac{1}{d}\right)^{2}}=\frac{1}{2\pi}K_{0}\left(\frac{|{\bf r}_{\perp}|}{d}\right)\ ,\label{int4EM}
\end{equation}
thus obtaining 
\begin{equation}
E^{II}=-\frac{qI}{2\pi}\frac{d^{3}d^{0}}{d\sqrt{(d^{0})^{2}-(d^{3})^{2}}}K_{0}\left(\frac{|{\bf r}_{\perp}|}{d}\right)\ ,\label{Ener12EM}
\end{equation}
where $K_{0}$ is a modified Bessel function of the second kind\,\cite{Arfken},
and 
\begin{equation}
|{\bf r}_{\perp}|=\sqrt{{\bf a}_{\perp}^{2}-\frac{({\bf d}_{\perp}\cdot{\bf a}_{\perp})^{2}}{(d^{0})^{2}-(d^{3})^{2}}}\ .\label{modb1EMM}
\end{equation}

Defining ${\hat{I}}$ as the unit vector along the straight line current
and noticing that $d^{3}$ is the projection of the vector ${\bf d}$
along ${\hat{I}}$, one can write the energy\,(\ref{Ener12EM}) in
the form 
\begin{equation}
E^{II}=-\frac{Iq}{2\pi}\ \frac{({\bf d}\cdot{\hat{I}})\ d^{0}}{d\sqrt{(d^{0})^{2}-({\bf d}\cdot{\hat{I}})^{2}}}\ K_{0}\left(\frac{|{\bf r}_{\perp}|}{d}\right)\ .\label{Ener14EM}
\end{equation}
This interaction energy is an effect due solely to the Lorentz violating
background, having no counterpart in Maxwell theory. Clearly, if the
background four-vector $d^{\mu}$ is zero, there is no interaction
energy. The energy (\ref{Ener14EM}) is proportional to the electric
charge $q$ as well as to the projection of the Lorentz-symmetry breaking
vector ${\bf d}$ along the current line. If the current line flows
perpendicular to ${\bf d}$, there is no interaction; the same happens
if ${\bf d}=0$. In the limit of a light-like background vector, $d\to0$,
the energy (\ref{Ener14EM}) vanishes.

The force on the point charge can be obtained from Eq.\,(\ref{Ener14EM})
as follows, 
\begin{align}
{\bf F}^{II} & =-\nabla_{{\bf a}_{\perp}}E^{II}\nonumber \\
 & =-\frac{qI}{2\pi|{\bf b}_{\perp}|}\ \frac{({\bf d}\cdot{\hat{I}})\ d^{0}}{d^{2}\sqrt{(d^{0})^{2}-({\bf d}\cdot{\hat{I}})^{2}}}K_{1}\left(\frac{|{\bf r}_{\perp}|}{d}\right)\nonumber \\
 & \ \ \times\Biggl[{\bf a}_{\perp}-\frac{({\bf d}_{\perp}\cdot{\bf a}_{\perp})}{(d^{0})^{2}-({\bf d}\cdot{\hat{I}})^{2}}{\bf d}_{\perp}\Biggr].
\end{align}
From Eq.\,(\ref{Ener14EM}), one can also obtain a torque on the
line current, due to the interaction with the point charge. Denoting
by $\phi$ the angle between ${\bf d}_{\perp}$ and ${\bf a}_{\perp}$,
we have 
\begin{align}
\tau^{II} & =-\frac{\partial E^{II}}{\partial\phi}\nonumber \\
 & =-\frac{qI}{4\pi g(\phi)}\ \frac{d^{0}\ {\bf d}_{\perp}^{2}\ ({\bf d}\cdot{\hat{I}})}{d^{2}\left[(d^{0})^{2}-({\bf d}\cdot{\hat{I}})^{2}\right]^{3/2}}\nonumber \\
 & \ \times K_{1}\left(\frac{g(\phi)}{d}\right)\sin2\phi\ ,\label{torquelc}
\end{align}
where we defined the function 
\begin{equation}
g(\phi)=\sqrt{{\bf a}_{\perp}^{2}-\frac{{\bf d}_{\perp}^{2}{\bf a}_{\perp}^{2}\cos^{2}\phi}{(d^{0})^{2}-({\bf d}\cdot{\hat{I}})^{2}}}\ .\label{modb1EMMM}
\end{equation}
If $\phi=0,\pi/2,\pi$, $d^{\mu}=0$ or ${\bf d}=0$, the torque in
Eq.\,(\ref{torquelc}) vanishes.

\section{\label{V}Dirac strings}

In this section we study the interaction between electromagnetic sources,
including Dirac strings. They might be seen as zero width solenoids
that connect magnetic monopoles, and their existence is compatible
with the standard Maxwell's electrodynamics, where they lead to the
Dirac quantization rule for the electric charge. In Maxwell electrodynamics,
a Dirac string does not produce any obvious physical effects, because
it does not produce electromagnetic field in its exterior region,
just a (divergent) magnetic field along the string. It is still relevant
to investigate whether a Dirac string can produce observable effects
in an extended electrodynamic theory. We will show this is indeed
the case, since we will show that non trivial interactions between
Dirac strings themselves and other electromagnetic sources will appear,
due to the presence of Lorentz symmetry breaking.

We start by considering a system composed by a point-like charge placed
at position ${\bf a}$ and a Dirac string, both of them stationary.
This system is described by the source 
\begin{eqnarray}
J_{\mu}^{III}\left({\bf x}\right)=J_{(D)\mu}\left({\bf x}\right)+q\eta_{\mu}^{0}\delta^{3}({\bf x}-{\bf a})\ ,\label{Dcurrent1}
\end{eqnarray}
where $J_{(D)}^{\mu}\left(x\right)$ stands for the source corresponding
to the Dirac string. Choosing a coordinate system where the Dirac
string lies along the $z$-axis with internal magnetic flux $\Phi$,
$J_{(D)}^{\mu}\left(x\right)$ is given explicitly by\,\cite{Fontes,FernandaDissertacao,AndersonDissertacao}
\begin{equation}
J_{(D)}^{\mu}({\bf x})=i\Phi(2\pi)^{2}\int\frac{d^{4}p}{(2\pi)^{4}}\delta(p^{0})\delta(p^{3})\varepsilon_{\ \ \nu3}^{0\mu}\ p^{\nu}e^{-ipx}\ ,\label{Dircurr2}
\end{equation}
where $\varepsilon^{\alpha\beta\mu\nu}$ is the Levi-Civita tensor
with $\varepsilon^{0123}=1$. If $\Phi>0$ we have the internal magnetic
pointing at the positive direction of $\hat{z}$, whereas for $\Phi<0$,
the internal magnetic field points in the opposite direction. In Maxwell
electrodynamics, the source\,(\ref{Dircurr2}) produces the vector
potential
\begin{equation}
A^{\mu}(x)=\frac{\Phi}{2\pi}\biggl(0,-\frac{x^{2}}{(x^{1})^{2}+(x^{2})^{2}},\frac{x^{1}}{(x^{1})^{2}+(x^{2})^{2}},0\biggr)\ ,
\end{equation}
which is, in fact, the vector potential related to a Dirac string,
with internal magnetic flux $\Phi$, lying along the $z$ axis.

From now on, the sub-index $\perp$ means the component of a given
vector perpendicular to the Dirac string. By following the same steps
presented in the previous sections, we obtain for the interaction
energy between the Dirac string and the point-like charge the expression
\begin{align}
E^{III}= & -i\frac{q\Phi d^{0}}{d^{2}}\int\frac{d^{2}{\bf p}_{\perp}}{(2\pi)^{2}}\frac{\left[\hat{z}\cdot\left({\bf p}_{\perp}\times{\bf d}_{\perp}\right)\right]}{\left({\bf p}_{\perp}^{2}+\frac{\left({\bf d}_{\perp}\cdot{\bf p}_{\perp}\right)^{2}}{d^{2}}\right)+\left(\frac{1}{d}\right)^{2}}\exp\left(i{\bf p}_{\perp}\cdot{\bf a}_{\perp}\right)\ .\label{EnerD15EM}
\end{align}
This integral can be manipulated similarly to Eq.\,(\ref{Ener10EMM}).
Using\,(\ref{defr}), we arrive at 
\begin{align}
E^{III}= & -\frac{q\Phi d^{0}}{d\sqrt{(d^{0})^{2}-(d^{3})^{2}}}\left[\hat{z}\cdot\left({\bf \nabla}_{{\bf r}_{\perp}}\times{\bf d}_{\perp}\right)\right]\nonumber \\
 & \times\int\frac{d^{2}{\bf q}_{\perp}}{(2\pi)^{2}}\frac{\exp\left(i{\bf q}_{\perp}\cdot{\bf r}_{\perp}\right)}{{\bf q}_{\perp}^{2}+\left(\frac{1}{d}\right)^{2}}\ ,
\end{align}
where ${\bf \nabla}_{{\bf r}_{\perp}}=\left(\frac{\partial}{\partial r^{1}},\frac{\partial}{\partial r^{2}},0\right)$.
After some manipulations, and identifying $\hat{z}={\hat{B}}_{int}$
as the unit vector pointing along the internal magnetic field, we
obtain 
\begin{align}
E^{III}= & \frac{q\Phi}{2\pi|{\bf r}_{\perp}|}\ \frac{d^{0}}{d^{2}\sqrt{(d^{0})^{2}-({\bf d}\cdot{\hat{B}}_{int})^{2}}}\ K_{1}\left(\frac{|{\bf r}_{\perp}|}{d}\right)\nonumber \\
 & \times\left[{\hat{B}}_{int}\cdot\left({\bf a}_{\perp}\times{\bf d}_{\perp}\right)\right]\ .
\end{align}
This interaction energy can be seen to lead to a force between the
Dirac string and the charge, as well as to a torque on the Dirac string.

The next example is given by two parallel Dirac strings placed a distance
${\bf a}_{\perp}$ apart. We take a coordinate system where the first
string lies along the $z$ axis, with internal magnetic flux $\Phi_{1}$,
and the second string lies along the line that crosses the $xy$ place
at ${\bf a}_{\perp}=(a^{1},a^{2},0)$, with internal magnetic flux
$\Phi_{2}$. The corresponding source is given by 
\begin{equation}
J_{\mu}^{IV}\left({\bf x}\right)=J_{\mu(D,1)}\left({\bf x}\right)+J_{\mu(D,2)}\left({\bf x}\right)\ ,\label{duasDcurrent1}
\end{equation}
where $J_{(D,1)}^{\mu}\left({\bf x}\right)$ is given by the right
hand side of Eq.\,(\ref{Dircurr2}), with $\Phi$ replaced by $\Phi_{1}$,
and 
\begin{equation}
J_{(D,2)}^{\mu}\left({\bf x}\right)=i\Phi_{2}\int\frac{d^{4}p}{(2\pi)^{2}}\delta(p^{0})\delta(p^{3})\varepsilon_{\ \ \nu3}^{0\mu}\ p^{\nu}e^{-ipx}e^{-i{\bf p}_{\perp}\cdot{\bf a}_{\perp}}\ .
\end{equation}
Proceeding as in the previous cases, and identifying the length of
the Dirac string as $L=\int dx^{3}$, we can show that the interaction
energy between the two Dirac strings is given by 
\begin{align}
E^{IV} & ={\Phi}_{1}{\Phi}_{2}L\Biggl[-\int\frac{d^{2}{\bf p}_{\perp}}{(2\pi)^{2}}e^{i{\bf p}_{\perp}\cdot{\bf a}_{\perp}}\nonumber \\
 & +\frac{1}{d\sqrt{(d^{0})^{2}-({\bf d}\cdot{\hat{B}}_{int})^{2}}}\int\frac{d^{2}{\bf q}_{\perp}}{(2\pi)^{2}}\frac{\left[\left({\bf d}_{\perp}\cdot{\bf q}_{\perp}\right)^{2}-{\bf q}_{\perp}^{2}{\bf d}_{\perp}^{2}\right]}{{\bf q}_{\perp}^{2}+\left(\frac{1}{d}\right)^{2}}e^{i{\bf q}_{\perp}\cdot{\bf r}_{\perp}}\Biggr]\thinspace.\label{EnerD7EM}
\end{align}
Provided that ${\bf a}_{\perp}$ is non-zero, the first term inside
the brackets in this result vanishes. The remaining integral can be
calculated with the procedure outlined in the previous sections. The
interaction energy between the two parallel Dirac strings per unit
length ${\cal {E}}^{IV}$ ends up given by, 
\begin{align}
{\cal {E}}^{IV}=\frac{E^{IV}}{L} & =\frac{{\Phi}_{1}{\Phi}_{2}}{4\pi}\frac{1}{d^{3}\sqrt{(d^{0})^{2}-({\bf d}\cdot{\hat{B}}_{int})^{2}}}\Biggl\{{\bf d}_{\perp}^{2}K_{0}\left(\frac{|{\bf r}_{\perp}|}{d}\right)\nonumber \\
 & +\frac{1}{{\bf r}_{\perp}^{2}}K_{2}\left(\frac{|{\bf r}_{\perp}|}{d}\right)\Biggl[{\bf d}_{\perp}^{2}{\bf a}_{\perp}^{2}\nonumber \\
 & -\Biggl(2-\frac{{\bf d}_{\perp}^{2}}{(d^{0})^{2}-({\bf d}\cdot{\hat{B}}_{int})^{2}}\Biggr)\left({\bf d}_{\perp}\cdot{\bf a}_{\perp}\right)^{2}\Biggr]\Biggr\}\ .\label{EnerD11EM}
\end{align}

The last example we consider is given by a Dirac string alongside
a steady line current, both parallel to each other. The corresponding
external source is 
\begin{equation}
J_{\mu}^{V}({\bf x})=I\eta_{\ \mu}^{3}\delta^{2}\left({\bf x}_{\perp}-{\bf a}_{\perp}\right)+J_{(D)}^{\mu}\left({\bf x}\right)\thinspace,
\end{equation}
where $J_{(D)}^{\mu}\left({\bf x}\right)$ is given by (\ref{Dircurr2}).
Proceeding as before, we obtain the result 
\begin{align}
{\cal {E}}^{V}=\frac{E^{V}}{L} & =\frac{I\Phi}{2\pi|{\bf r}_{\perp}|}\ \frac{{\bf d}\cdot{\hat{B}}_{int}}{d^{2}\sqrt{(d^{0})^{2}-({\bf d}\cdot{\hat{B}}_{int})^{2}}}\nonumber \\
 & \times K_{1}\left(\frac{|{\bf r}_{\perp}|}{d}\right)\left[{\hat{B}}_{int}\cdot\left({\bf a}_{\perp}\times{\bf d}_{\perp}\right)\right]\ ,
\end{align}
for the energy line density.

The results of this section are all exclusive effects of the LV, having
no counterpart in Maxwell theory, in which the interaction energy
vanishes in all cases considered here. In the limit $d^{\mu}\rightarrow0$,
all these effects disappears, as they should. The same happens in
the limit of a light-like background vector, $d\to0$.

\section{\label{conclusoes}Conclusions and perspectives}

In this paper we investigated the interaction between sources for
the electromagnetic field in the presence of the Lorentz violating
higher derivative interaction $d^{\lambda}d_{\alpha}\partial_{\mu}F_{\nu\lambda}\partial^{\nu}F^{\mu\alpha}$.
This interaction is induced by a specific setting of nonminimal LV
which was shown to lead to low energy effects relevant for the physics
of light pseudoscalars interacting with photons\,\cite{ALP,Highoperators}.
We obtained results with no resource to perturbation theory in the
background vector for the specific case where $d^{\mu}d_{\mu}=d^{2}>0$
(time-like interval), which provided us with different physical effects
with no counterpart in Maxwell theory. The case of a light-like background
vector, $d\to0$, can be obtained from our results. In this situation,
the interaction between two point-like charges becomes the Coulombian
one, and all other nontrivial interactions obtained in this paper
vanish. On the other hand, a space-like background vector was not
considered here. In this situation the calculations are much more
difficult and some preliminary results suggest that it could lead
to inconsistencies.

We have shown the emergence of an spontaneous torque on a classical
electromagnetic dipole, as well as a nontrivial interaction between
a steady straight line current and a point-like charge. We also investigated
some phenomena due to the presence of Dirac strings. We showed that
a Dirac string have a nontrivial interaction with a point charge,
with a straight line steady current, as well as with another Dirac
string. All these phenomena are effects due to the Lorentz-violation
background. The nontrivial LV effects uncovered in this paper represent
another instance where nominimal LV terms may induce low energy phenomenology,
and might open up a window to look for experimental limits on these
LV coefficients.

As a final remark, we point out that this paper all field sources
are spinless. An interesting extension of this work would be the investigation
of spin effects in the interactions between field sources.

\bigskip{}

\textbf{Acknowledgments}

This work was supported by Conselho Nacional de Desenvolvimento Científico
e Tecnológico (CNPq) and Fundação de Amparo a Pesquisa do Estado de
São Paulo (FAPESP), via the following grants: CNPq 482874/2013-9,
FAPESP 2013/22079-8 and 2014/24672-0 (AFF), FAPESP 2013/01231-6 (LHCB),
CNPq 484736/2012-4 and 311514/2015-4 (FAB).

\end{document}